___

# Limb Event Brightenings and Fast Ejection Using IRIS Mission Observations.

E. Tavabi[1] . S. Koutchmy[2] . L. Golub[3]

**Abstract** The *Interface Region Imaging Spectrograph* (IRIS) of the recently commissioned NASA small explorer mission provides significantly more complete and higher resolution spectral coverage of the dynamical conditions inside the chromosphere and transition region (TR) than has heretofore been available. Near the solar limb high temporal, spatial (0.3") and spectral resolution observations from the ultraviolet IRIS spectra reveal high-energy limb event brightenings (LEBs) at low chromospheric heights, around 1 Mm above the limb. They can be characterized as explosive events producing jets. We selected two events showing spectra of a confined eruption just off or near the quiet Sun limb, the jet part showing obvious moving material with short duration large Doppler shifts in three directions identified as macrospicules on slit-jaw (SJ) images in Si IV and He II 304 Å. The events are analyzed from a sequence of very close rasters taken near the central meridian and the South Pole limb. The processed SJ images and the simultaneously observed fast spectral sequences having large Doppler shifts, with a pair of red shifted elements together with a faster blue shifted element from almost the same position, are analyzed. Shifts correspond to velocities of up to 100 km s$^{-1}$ in projection on the plane of the sky. The occurrence of erupting spicules and macrospicules from these regions is noticed from images taken before and after the spectra. The cool low first ionization potential (FIP) element simultaneous line emissions of the Mg II h and k resonance lines do not clearly show a similar signature due to optical thickness effects but the Si IV broad-band SJ images do. The bidirectional plasma jets ejected from a small reconnection site are interpreted as the result of coronal loop-loop interactions leading to reconnection in nearby sites.

[1] Physics Department, Payame Noor University (PNU), 19395-3697-Tehran, I. R. of Iran, Email: tavabi@iap.fr

[2] Institut d'Astrophysique de Paris, UMR 7095, CNRS and UPMC, 98 Bis Bd. Arago, 75014 Paris, France,

[3] Harvard-Smithsonian Center for Astrophysics, 60 Garden St., Cambridge MA 02138, USA.

## 1. Introduction

In their seminal paper Brueckner and Bartoe (1983) described explosive events (EEs) from on-disk far-ultraviolet (far-UV) observations in hot transition region emission lines with formation temperatures ranging from logT=4.3 to 5.7. These temperatures are similar to those more extensively studied with the new *Interface Region Imaging Spectrograph* (IRIS) mission observations that we will discuss in this article, see Figure 1 for a comparison. They identified on-disk EEs, defined before in Brueckner, 1980 as: i) features showing high turbulence confined in small areas showing a very broad line-profile and ii) jets showing upward -moving material with velocities exceeding the sound velocity (~120 km s$^{-1}$) in the corona, see Figure 1. In addition their sequence of C$_{IV}$ spectroheliograms confirmed the occurrence, above the polar region limb, of spikes and jets similar to spicules and macrospicules first discovered from the *Skylab* mission data. EEs were hypothesized to explode from small loops with a size of 3000 km or less. Dere, Bartoe, and Brueckner (1989) from the same *High Resolution Telescope and Spectrograph* (HRTS) rocket mission observations, analyzed chromospheric jets that identified. FUV and EUV spectroheliograms from the *Skylab* mission showed small surge like eruptions, called macrospicules (because of their similarity to ordinary Hα spicules), on the limb in coronal holes, especially in the polar regions. Bohlin *et al.* (1975) established the ranges of size, shapes, and other physical parameters for these features. Macrospicules were again observed with the Spacelab 2 mission. Moore *et al.* (1977) presented observational evidence of a connection between macrospicules and X-ray bright point flares and Golub *et al.* (1974) found a close relationship with X-ray bright points (observed in soft X-ray filtergrams). Macrospicules are sudden eruptive events usually having a rise time of a few minutes or less. If spicules are just small-scale versions of macrospicules, the relationship of macrospicules with X-ray bright point flares could suggest that spicules are produced by similar mechanisms at smaller scale and lower temperatures, but this is far from being established by observations because the hot counterpart of spicules has never been clearly observed. Indeed from a unique rocket observation Daw, DeLuca, and Golub (1995) convincingly showed that the bulk of spicules is recorded only in absorption against the coronal off-limb background. In polar regions, some evidence of association between hot "jetlets" and tiny transient soft-X ray (SXR) brightenings was seen from *Yohkoh* mission partial frame sequences (Koutchmy *et al.*, 1997).

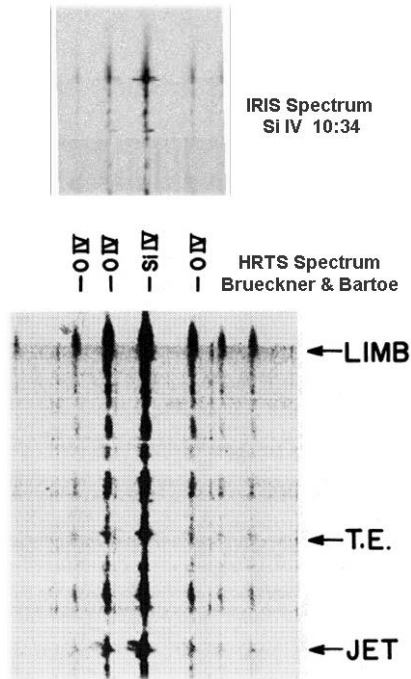

**Figure 1:** Negative snapshots selected to show the limb and on-disk line emissions near a polar region. The bottom panel is a partial frame from the classical HRTS photographic observations. The upper panel is an IRIS partial frame taken at very high temporal resolution to show a LEB in the Si IV line discussed in the text. Spectra are aligned and scales are similar but the HRTS spectrum is exposed longer and is taken with a larger slit width, giving a lower resolution but a lower threshold.

From eclipse observations, Koutchmy and Stellmacher (1976) found very thin (~1") white-light coronal spikes reaching coronal temperatures above the polar chromosphere. They are visible due to Thomson scattering produced at 5 to 20 Mm heights; their densities are of the order of the jet densities ($1 \times 10^{10}\,\text{cm}^{-3}$) and the altitude of the spikes is similar to the largest jets reported by Brueckner (1980) seen in C IV (logT=5) spectroheliograms (Brueckner and Bartoe 1983).
Off-limb macrospicules and surges in polar regions were observed with ground-based telescopes in Hα and other low-excitation lines. With *the R.B. Dunn Solar Telescope and the Universal Birefringent*

*Filter* (DST-UBF) of the National Solar Observatory (NSO) at Sacramento Peak a comparison was made using helium He$_{II}$ 304 Å images from the *Extreme ultraviolet Imaging Telescope* (EIT) of the *Solar and Heliospheric Observatory* (SoHO) by Georgakilas, Koutchmy, and Alissandrakis (1999), including the description of complex erupting polar region structures with large amplitude motions. Small radial jets, possibly related to Hα polar limb surges (Koutchmy and Loucif, 1991, Georgakilas, Koutchmy, and Alissandrakis 1999), are particularly prominent in polar coronal holes where transient SXR brightenings were observed from *Yohkoh* mission images (Koutchmy et al., 1997). These are the close neighbors of tiny bright loops seen in the Ca II H images of the solar optical telescope (SOT) of the *Hinode* mission in polar regions (Tavabi, Koutchmy, and Ajabshirizadeh 2011 and 2015). The geometric shape of the jets and their locations suggest that they arise near singular null points of the coronal magnetic field. On disk, the large number of events observed near or outside plages, coupled with the high velocities of the apparent outflows, indicates that the jets may contribute to the high-speed solar wind (Cirtain *et al.,* 2007).

Magnetic reconnection has been suggested as a possible mechanism for generating spicules or jet-like features (*e.g.* Yokoyama and Shibata, 1995; Moore *et al.,* 2010). We note that regarding spicules, they are classically observed in: i) cool low excitation lines like Hα and Hβ of H I or in the He I lines, that are high first ionization potential (high FIP) lines above 10 eV and ii) low FIP lines, like the powerful resonance lines of Ca$_{II}$ in the UV and in the infrared (IR). Sterling (2000) and more recently Tsiropoula *et al.* (2012) presented reviews of spicule morphological properties, including their derived physical parameters and dynamical behavior. Earlier Pasachoff, Jacobson, and Sterling (2009) showed that spicules appeared to have bidirectional plasma flows at their maximum heights consistent with the magnetic reconnection scenario introduced by Innes (1997) from much larger scale off limb spectrograms taken in hot transition region (TR) lines *by the Solar* Ultraviolet Measurements of Emitted Radiation (SUMER) instrument onboard the SOHO. Tsiropoula, Alissandrakis, and Schmieder (1994) reported a similar behavior in mottles on-disk. Chromospheric bright points have been observed in cases where there are emerging arch-shaped mottles, and thus the bright points result from reconnection (Singh *et al.,* 2012), where a current sheet provides a strong acceleration of charged particles as is the case for larger scale events (Filippov, Koutchmy, and Tavabi 2013). Emerging miniature arches in the chromosphere appear to play a key role in many X null-point reconnection sites (Shimojo et al., 1996). Cirtain *et al.* (2007) and Filippov, koutchmy, and Golub (2009b) reported plasma jets of various scales from *X-Ray Telescope* (XRT) observations of the *Hinode* mission, from giant X-ray jets to numerous small jets with sizes typical of macrospicules. Macrospicules are now continually observed using the 304 Å resonance line emission of HeII from full disk imaging experiments such as the *Atmospheric Imaging Assembly* (AIA) of the *Solar Dynamics Observatory* (SDO) mission.

It seems clear that near limb and even better, off- limb higher resolution observations, would permit a better characterization of all brightening events, as far as the height parameter is concerned, in case of observations with high enough resolution to resolve the reconnection regions involved. We believe the new IRIS observations will let us tackle this point.

## 2. Observations and Data Reduction

IRIS 0.3"-0.4" spatial resolution observations, with a pixel size of 0.166" (corresponding to 120 km at disk center) and high cadence, reveal the dynamical behavior of the chromosphere and of TR fine features (De Pontieu *et al.,* 2014). The available high cadence allows the analysis of the evolution of these features and enables us to follow them from the photosphere to the low corona in channels (Figure 3) of different spectral lines, in the near ultraviolet (NUV, 2782-2835 Å with the MgII h and k resonance lines seen in absorption on-disk and in emission off-disk, similarly to the well-known H and K lines of CaII), and in the far ultraviolet (FUV, 1331-1358 Å with the CII line seen in emission) and in 1389-1407 Å (with the SiIV lines seen in emission). Note that the MgII lines are due to a low-FIP element typical of the low temperature plasma produced above the temperature minimum near the 500 km height above $\tau_5 = 1$ (optical deep in 5000 Å). Their FIP is 7.65 eV. The CII line is from a high-FIP element with excitation potential corresponding to a higher temperature. The SiIV lines are produced at much higher temperatures with an ionization potential of $Si^{+2}$ of 33.5 eV. The velocity resolution in IRIS spectra is 1 km s$^{-1}$. In addition SJ images with 175×175 arcsec$^2$ field-of-view (FOV) reflected off the slit/prism assembly through a filter wheel with four different filters (De Pontieu *et al.,* 2014) are available.

Figure 2 shows a SJ with a faint roundish bright dot in both hot FUV channels (1400 Å and 1330 Å), near the center of the FOV and off limb, with the entrance slit of the spectrograph crossing the region at 10:34 UT. No clear evidence of anything exceptionally bright in the simultaneous NUV cool line (2796 Å) SJ exists; the lifetime of the phenomenon is of order 100 seconds from inspection of the SJ images.

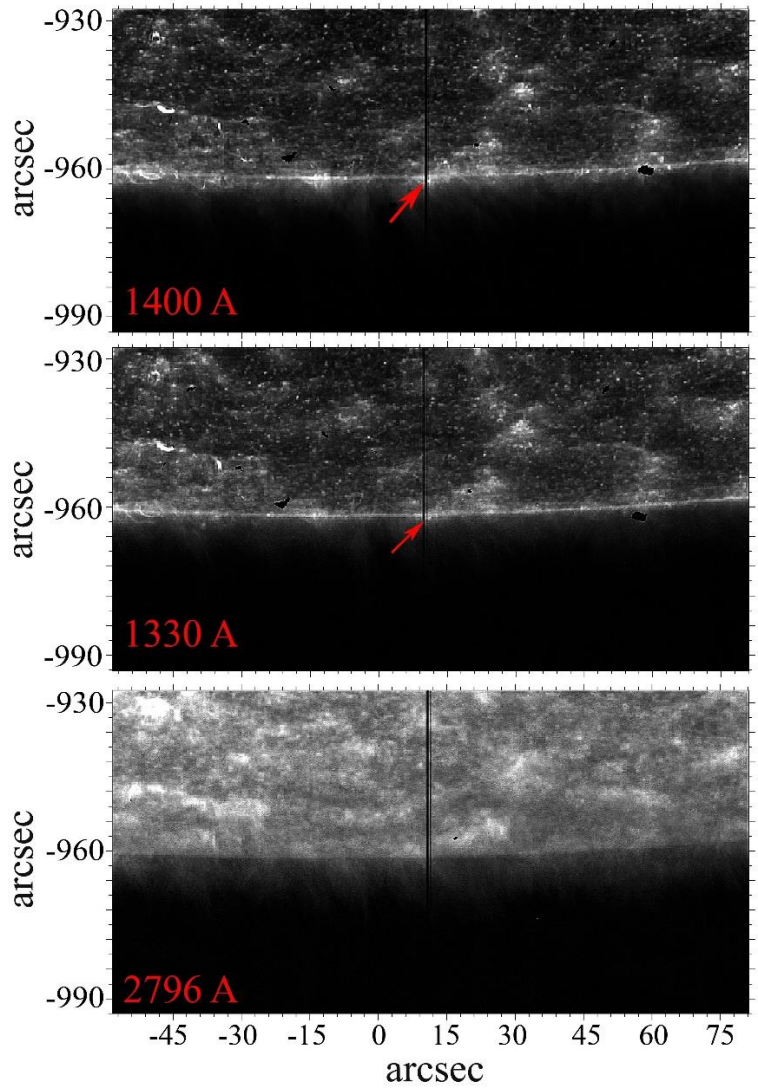

**Figure 2:** Snapshots of IRIS SJ for Si IV (logT~4.8), C II (logT~4.3) and Mg II k (logT~4.0) taken near the South Pole on 10 October 2013 at 10:34 UT. The time difference between these frames is of the order of 8.9 seconds. The arrow shows the bright event adjacent to the limb seen in the TR lines (Si IV; C II) right under the slit of the spectrograph, shown by the vertical black line. The spicule "forest" above the limb is apparent in all images, with rather good correlation between them. Note the apparent absence of the bright event in the SJ image in the Mg II cool line above the limb in this high-resolution image with minimal contribution from the adjacent continuum, the vertical and horizontal axes indicate the distance form Sun center.

Our region is exactly above a solar pole (Figure 3), which is usually a quiet region and no active region is observed. It is not a coronal hole region. Small patches of concentrated magnetic field can however exist in polar region, Figure 3 provides a synoptic image of the region seen in 304 Å He II emission from AIA observations during the IRIS observations reported here. A very large IRIS raster over the Sun's South pole was done on 10 October 2013 at 10:02 to 11:01 UT centered at [6", -945"].

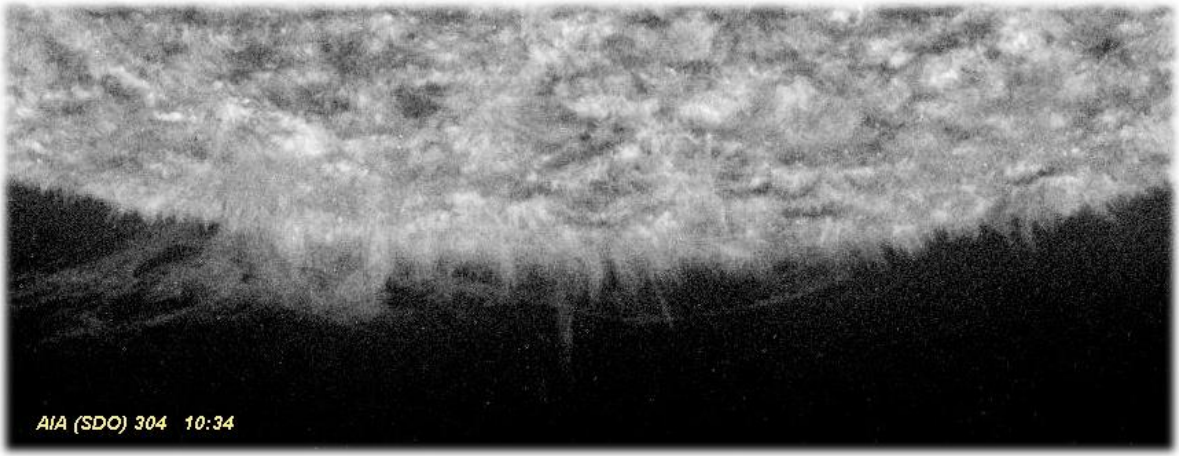

**Figure 3:** Partial frame taken with AIA showing the South polar region where IRIS spectroscopic and SJ near limb and off- limb observations are taken over a larger FOV. Snapshot taken at 10:34 UT on 10 October 2013 using the He II 304 Å emissions.

Calibrated level 2 data were used. The raster cadence is 9 seconds with a spatial step of 0.35". The 400-pixel long slit covers a 141x174 arcsec$^2$ region. The SJ high-resolution images were obtained in three channels, showing: i) the low temperature chromosphere in Mg II k (2796 Å, logT~3.6 to 4), and ii) the much hotter transition region in Si IV (lines near 1400 Å with logT~4.8) and iii) the C II low transition region filter (1330 Å, logT~4.0). Their FOV is 167×174 arcsec$^2$ with a 36 seconds cadence and pixel size is of the order of 0.1" (De Pontieu *et al.,* 2014).

Because we want to precisely look at LEBs and evaluate the height of the event with respect to the photospheric limb, it is important to use a reference showing the limb in spectra and to align precisely the spectra where the limb is not clearly seen (Si IV spectra). This can be done with reference to fiducial mark or hair line (hl) put on the entrance slit of the spectrograph although it is not so easy to find the signature of this hl in the spectra. A rather deep processing is needed, see Figure 4, where the method is illustrated using the example of a LEB taken at 10:34 UT that we believe clearly shows the LEB definitely above the photospheric limb seen on Mg II h and k line spectra.

### 3. Results

In classical observations taken above the limb, a bright background obscures or smears the signature of small events because of the line-of-sight (LOS) effects, *i.e.* a long integration path along the line of sight with superposition or overlap effects appearing in projection in the 2D images. Note that images taken near the disk center do not suffer as much from LOS effects. However, it is believed that TR structures are indeed of very small cross section, corresponding to a tiny value of the filling factor, $10^{-2}$ to $10^{-5}$ (Judge, 2000). Thanks to the very high spatial resolution of FUV observations performed with IRIS the opportunity to observe the tiniest events that can be found in the region of the quiet Sun for the first time without suffering too much from LOS effect.

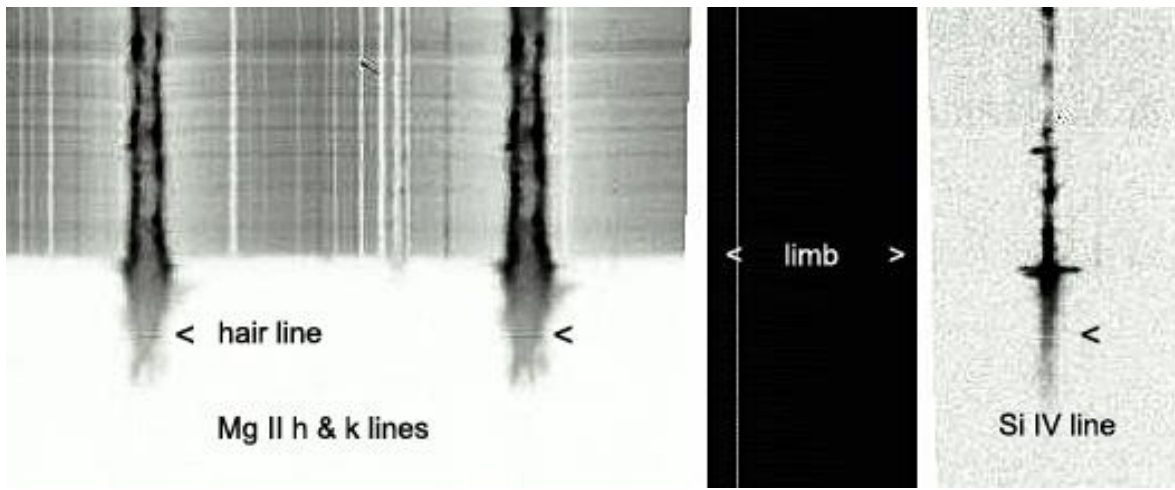

**Figure 4**: Composite prepared to illustrate the method applied to precisely put the reference limb level in FUV images taken near the SiIV lines at 10:34 UT at left negative image obtained near the UV MgII h and k lines where the limb is clearly apparent at left with the precisely aligned negative of EUV spectrum of SiIV with the position

of the limb shown in the black box, the signature of the hair line as a reference as shown with < in the left and right panels. Note the small on-disk event showing a large shift excursion in SiIV and a very small signature in the MgII line wings at almost the same disk position, confirming that the alignment is correct.

To give a quantitative evaluation of the dynamical bright event (flashing LEB) using spectra, the bright feature should be observed with the slit of the spectrograph crossing it (Figure 5). We emphasize that due to the short lifetime and the very small size of the bright knot, it is relatively rare so far to optimally observe such event, with the IRIS spectrograph slit pointed exactly "over" the event during the raster. Fortunately in our case, the slit crosses the limb *at the right* time when the bright dot flashes (see Figure 5). Doppler shifts can then be detected and measured. In case of a long and broader slit put on-disk, these events are easier to pick up as was done for the events reported by Brueckner and Bartoe (1983) but their fine structure could not be analyzed as shown on Figure 5. We evaluate the typical extension along the limb of this LEB to be three steps of the spectroscopic raster, or 1.05". From Figure 4 the typical height of the SiIV LEB above the limb is evaluated to be 4 pixels or 0.67" (see also Figure 7).

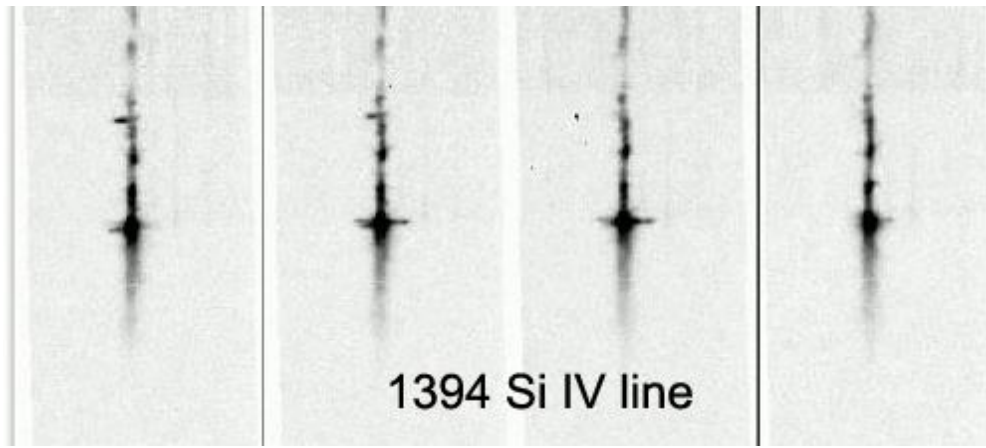

**Figure 5**: Snapshots of partial frame successive SiIV line spectra obtained at 10:34 UT when the slit crossed the LEB of Figure 4 with 9 seconds cadence with a spatial step of 0.35". Note the rapid changes in the line shape with blue and red extensions. On the disk several small events are also crossed. The position of the limb is evaluated as shown in Figure 4.

Figures 6 show Dopplergrams to illustrate the spatial distribution of the LOS velocity components associated with the feature.

Dopplergrams were constructed by subtracting red and blue wing intensities at the same wavelength offset from the rest, the bright-dark (positive-negative) corresponds to towards (blue shift)-red shift (towards-away from the observer). The position of the LEB is marked by the green arrows as well as the very rapid blue and co-temporal red shifts, in 1403 Å Si IV raster Dopplergrams in the velocity space. Both Si IV channels show similar behaviors, the morphological properties in all ranges of Doppler velocities are similar; in both channels we possibly see an elongated typical spicule with an inclination angle having a high velocity value (blue arrows). The brightest knot-like feature of the LEB is not seen in the Mg II k line (Figure 6, third row), while the thicker spicules (more like a macrospicule or a spicule bush) appear definitely *above* the location of the LEB (bright means blue shifts). In these figures we see fewer features in the far wings than near the center of the line (panels from left to right).

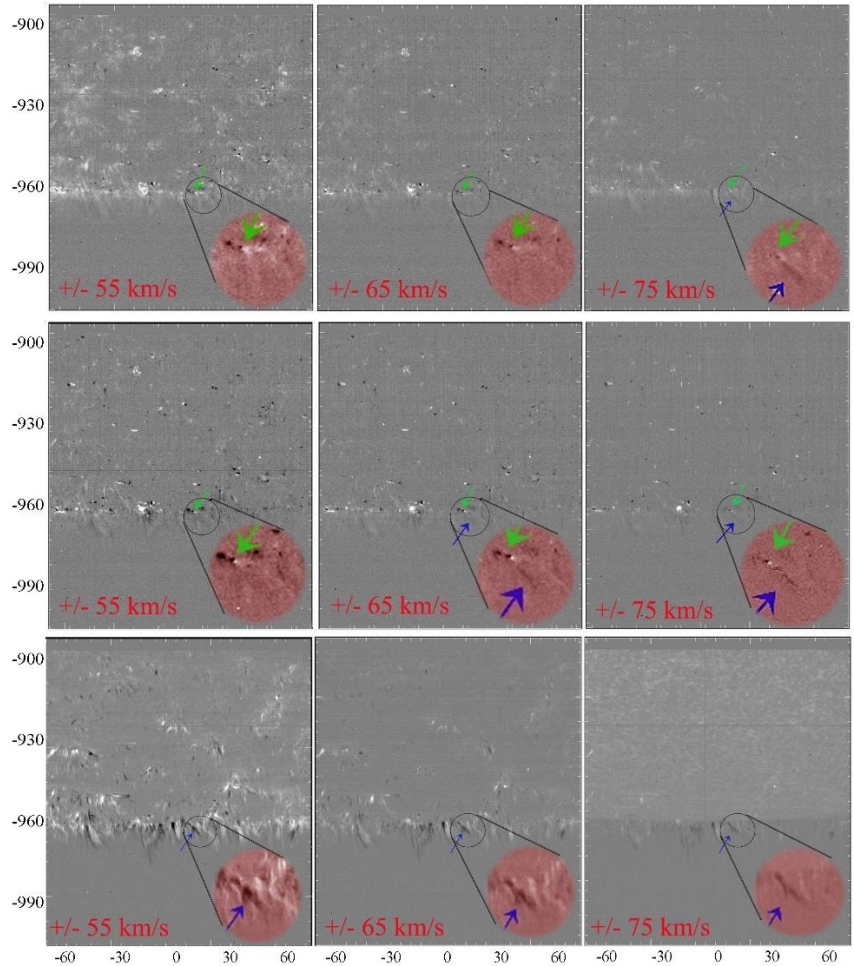

**Figure 6**: Filtered in the velocity space (white is for blue shifts and black for red shifts) for three different velocity ranges. First row is for SiIV (1403 Å), second raw for SiIV (1393 Å) and the last row is for MgII k (2796 Å), the position of the LEB is marked by the green arrows and the rapid ejection is shown by the blue arrows, the vertical and horizontal axes indicate the distance form Sun center in arcsec.

Figure 7 shows the central simultaneous spectrogram in two SiIV lines corresponding to the SJ images of the LEB knot (10:34 UT) shown in Figure 3 and the Dopplergrams of Figure 6. Both lines have strong similarities in their details, including asymmetries seen in the blue and the red wings. Unlike what is seen in these lines, the MgII k cooler line shows a broad doubly peaked shape typical for a large and broad limb made of spicules and possibly macrospicules seen along the LOS with significant optical thicknesses at this position, but no sign of the strongly Doppler shifted components that were seen in SiIV. The red wing of the hot lines shows a pair of shifted components of material with a definite and small separation distance between them (0.5" or 360 km). The shift is comparable in extension to the blue wing shift. The shifted components correspond to flows spanning the whole range of velocities from 0 to approximately 100 km s$^{-1}$ without taking into account the LOS effect that could drastically reduce the values of the apparent velocities in case propagation is occurring in a direction with a large departure from the direction of the LOS. It is not excluded that some large flows also exist in the direction corresponding to the image plane without producing any Doppler shifted component. We note that the line profile above the peak of maximum intensity is broad over approximately 2". The splitting of the two largely red shifted components is an important feature in case we deal with the so-called diffusion region where reconnections are taking place (Aschwanden 2014), the region extends over at least 500 km in height (see Figure 7).

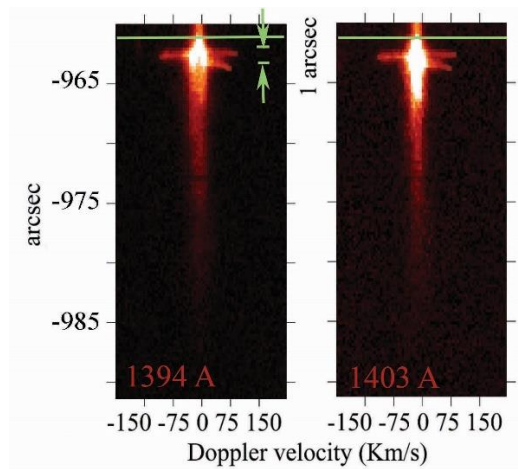

**Figure 7**. Highly contrasted and magnified simultaneous spectrograms of the South pole LEB in different Si IV lines taken on 10 October 2013 at 10:34 UT. In the direction perpendicular to the image plane, ejections are evidenced in these 1394 and 1403 Å Si IV lines by the large Doppler effects at positions shown near the 962" mark (scale at left) which is typically at a 1" distance from the limb (see also Figure 4). This position corresponds to the bright feature in figure 3, top two panels. Note the curved red wing extension of the upper jet, the vertical axis indicate the distance form Sun center.

Figure 8 shows the integrated line profile intensity variations at the position where the LEB appeared on the raster (approximately 1 Mm or less in projection above the visible limb). The maximum value of the intensity was about three times the background intensity in the Si IV hot channels (without any correction for smearing). The total lifetime of the brightening event is not well known but assuming the LEB is similar to the EEs of Brueckner (1980), it is about 80 seconds, in agreement with the observed behavior resulting from the analysis of the SJ images that show a 1.5 min lifetime for the LEB at 10:34 UT.

This figure shows what we interpret as a reconnection site because there is evidence of a sudden large brightening recorded in the hot lines only and that a non-thermal heating occurs in this region. The very narrow ejection of plasma seen as a long and tiny thread shown in the SJ images (top of Figure 10 and Figure 11) is another piece of evidence that we interpret in favor of a reconnection event. The absence of evidence in the cool lines finally also favors this interpretation because in the hot but much lower density plasma, the magnetic field plays a more important role.

Another possibility to have a sudden heating would be to consider a shock, but we do not have any evidence of a wave occurring in this region; only some violent ejection phenomena are observed.

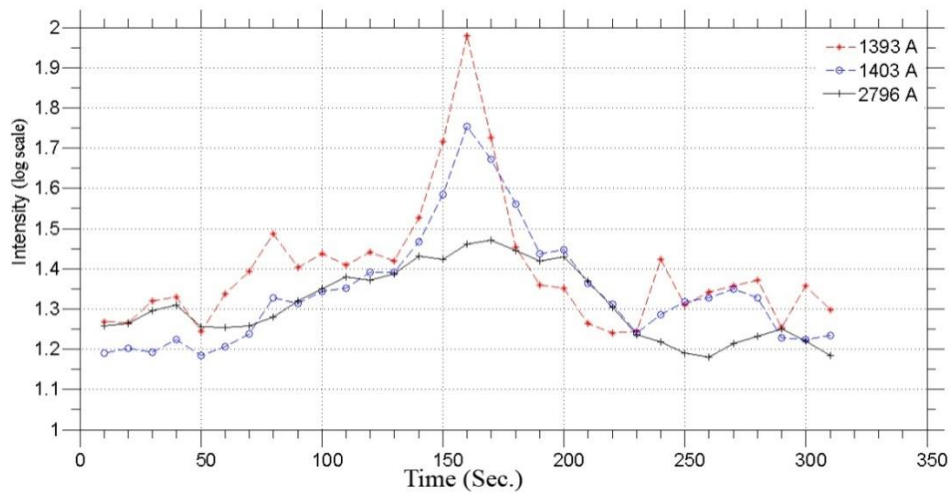

**Figure 8:** *Normalized intensity variations (logarithmic scale) for each selected spectral lines around the bright feature of 10:34 UT that we interpret as a reconnection site. The red curve reflects the extension in space of the reconnection region assuming no intrinsic time variations.*

Figure 9 (top) shows details of the evolution during the raster of the Si IV (1393 Å) line spectra profiles at the position coinciding with the LEB as shown in the SJ image in Figures 3 and 7 and marked there with an arrow. The bottom panel shows the difference between the line profiles and the spectrum averaged over the full length of the slit (Sekse, Rouppe van der Voort, and De Pontieu 2012). The difference spectral profiles show more clearly the "true" Doppler shifts, neglecting LOS effects. Strong line shifts appear on both sides of the zero velocity position after the third time step and reveal a short lived asymmetry in the blue wing of this broad emission line, that suddenly increases in both the blue and red directions, with a significant line of sight upward velocity component. Peak values in line profile are of order of 150 km s$^{-1}$ and they are (temporally correlated with) a red shift (downward motion) of around 100 km s$^{-1}$.

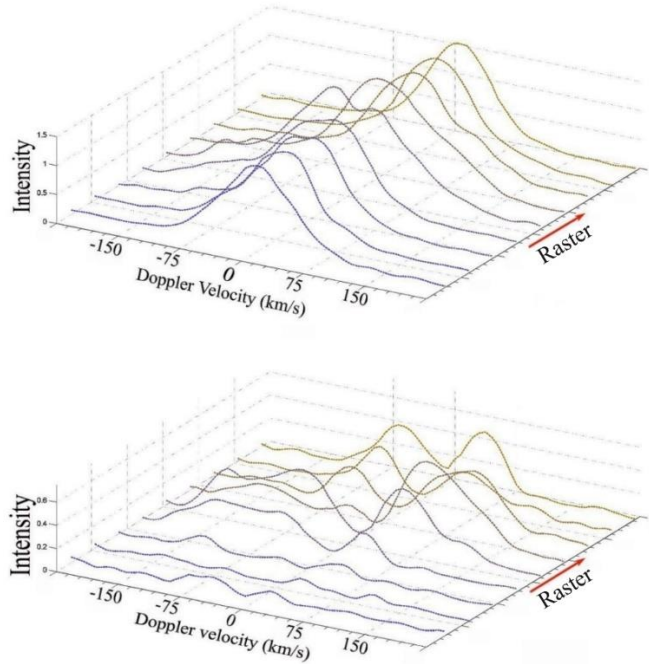

**Figure 9:** (top) Velocity/position diagram from the Si IV (1393 Å) line profile. (bottom) After subtracting the averaged value along the spectral line to obtain the relative Doppler red and blue "true" shifts, without taking into account LOS effects. The time interval corresponds to the first and last curves between 10:33:57 to 10:35:08 UT with a cadence of around 8 seconds and a shift in space of 0.35" between each spectrum.

The strong intensity enhancement in the hot lines of Si IV provides some evidence of heating; in addition Figure 9 suggests some dispersion of velocities inside the LEB. To improve the signal/noise ratio for off-limb parts of the SJ images, we decided to use a logarithmic scale (Figure 10) and process the images using an unsharp masking filter with a three pixle size window; an increased contrast was also used. The time evolution in Si IV is shown in Figure 11 with clear evidence of a thin long linear thread emerging from the bright knot region of the LEB. *After 2 minutes* the thread appears also noticeable in Mg II k line at the same position but with a thicker and shorter length (Figure 12).

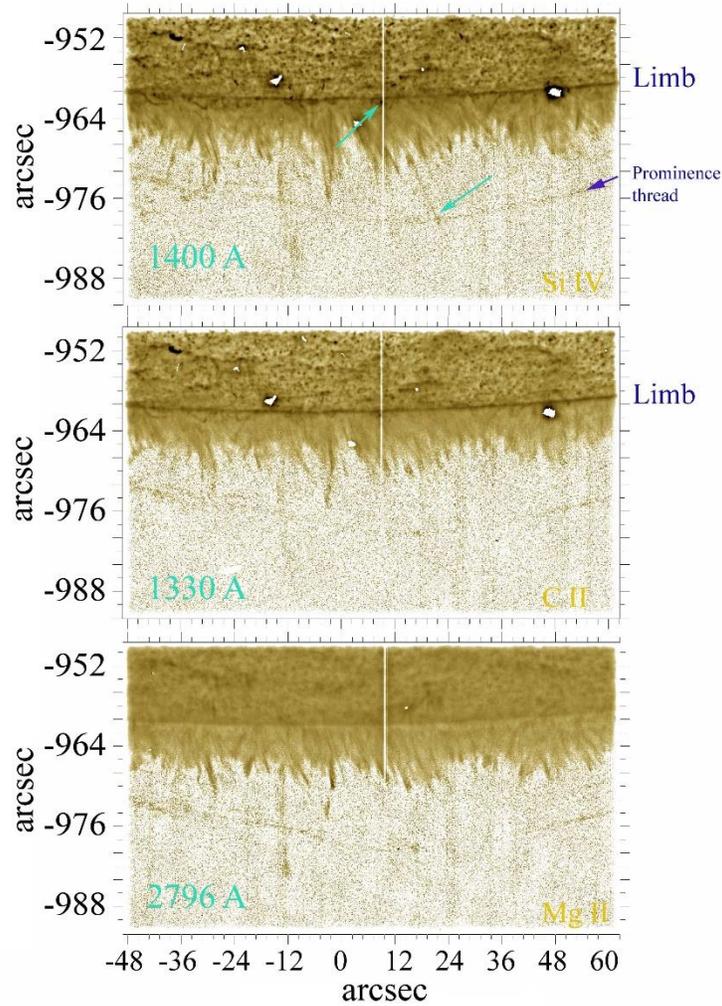

**Figure 10:** Highly processed SJ images in three wavelengths taken at 10:34 UT, the vertical and horizontal axes indicate the distance form Sun center. In the hottest line (SiIV 1400 Å, logT~4.8) a very sharp and dynamic thread (top and bottom are indicated by arrows) seems connected to the reconnection site, the lifetime being less than 1.5 minutes (see Figure 8).

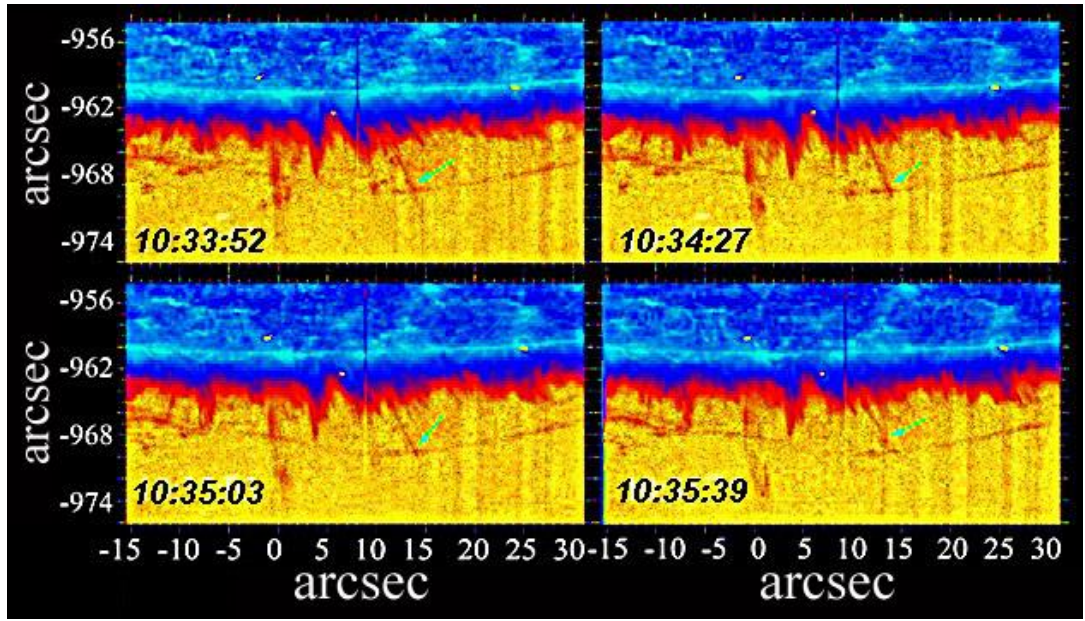

**Figure 11:** Time evolution of the tiny linear and short lifetime thread (green arrows) recorded above the site of the LEB and aligned with its direction at 10:33 UT, the vertical and horizontal axes indicate the distance form Sun center. Frames of the 1400 Å SJ channel during the crossing by the entrance slit of the spectrograph. The image intensity is scaled logarithmically and shown in reverse scale with color coding for making clearer the linear thread. Note also the quasi- horizontal thread (crossing the whole FOV) probably related to a polar region filament, see Figure 3.

The thin linear jet-like thread of Figure 11 seems to originate from the bright point that is marked by arrows in Figure 10 (top panel) that we called LEB. This thin linear long feature was not noticed in the original and unprocessed SJ images; it is seen only in the SJ hottest channel (1400 $Å$) after some image processing, see Figure 10. We hesitate to call the feature a macro- spicule; it is definitely longer and evidently hotter than the usual spicules that are shown using the cooler emissions of Mg_II, see Figure 12. It is noteworthy that the origin of this fast jet was probably detected in Dopplergrams of large amplitude velocities as a thicker blue shifted material appearing in Figure 6. Its counterpart in Figure 12 is not obvious although some feature(s) appear at the same location but later in time.

**IRIS Limb Event Brightenings and Fast Ejection**

_________________________________________________________________________________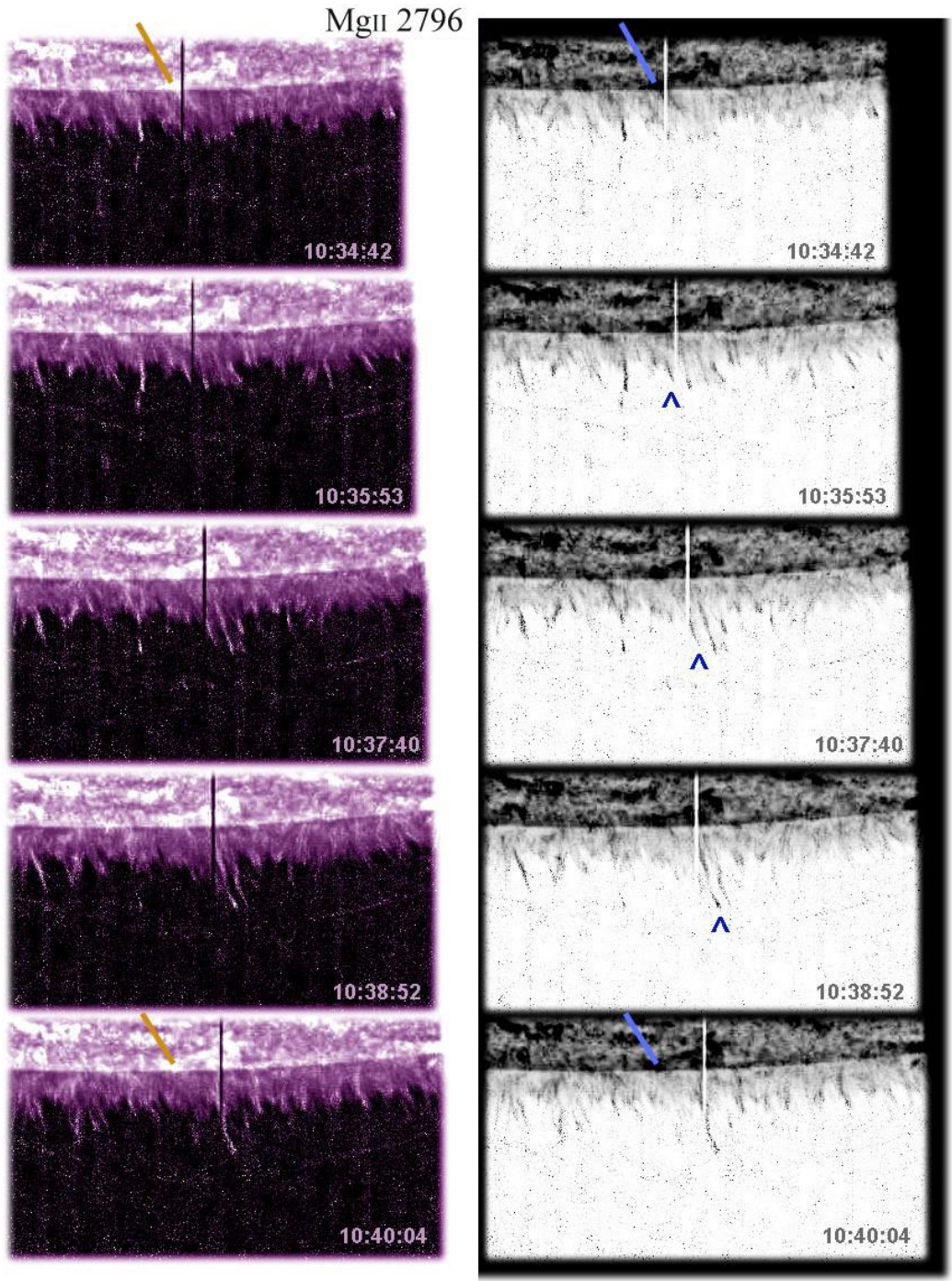

**Figure 12:** Selected successive images in both positive (at the left) and negative (at the right) displays to show the thicker and shorter "candidate" features that appeared in the rather cool MgII 2796 Å SJ images at positions that correspond to the thread in Figure 11. The LEB is shown by a bar to the left of the slit on-disk (compare with Figures 10 and 11). Note the time inserted and the displacement of the entrance slit of the spectrograph during the raster.

### 4. Discussion

The most significant aspect of spicules and especially of macrospicules is perhaps their potential ability to transfer mass and energy from the photosphere to the chromosphere and to the bottom of the corona. In a precursor study of Transition Region And Coronal Explorer *(*TRACE) Lyα and quasi- simultaneous CIV broad band filtergrams, Alissandrakis, Zachariadis, and Gontikakis (2005) found a lot of similarity between spicules recorded in the rather cool but optically thick Lyα line and in the hotter CIV line (both are high-FIP elements and the CIV line behaves similarly to our SiIV lines).

The bright points at the feet of jets (Tsiropoula and Tziotziou 2004; Sterling, Moore, and DeForest 2010; Tavabi 2014 and Suematsu *et al.,* 1995) have been seen in many chromospheric and coronal emission lines (Moore *et al.,* 2010) and often interpreted as the result of magnetic reconnection, where the magnetic free energy is suddenly and partially converted to thermal and kinetic energy in the region of the null-point with much weaker magnetic field strength. Some of the magnetic field lines then open up and the hot plasma is readily able to move along these open field lines outwardly, being quickly accelerated. The gas pressure gradient works like a piston that pushes plasma through a nozzle (Filippov, Koutchmy, and Tavabi 2013). Emission lines exhibit Doppler shifts towards red and blue as one would expect from a bidirectional jet (Innes *et al.,* 1997) at TR temperatures.

In our LEB of 10:34 UT we saw evidence of a significant extension of the reconnection region, of order of 400 km in height and 800 km in azimuth (see Figures 4, 5 and 7). This is also suggested in Figure 8 where the FWHM of the LEB in SiIV lines extends over approximately four positions of the raster, which corresponds to a 1" distance along the limb. At lower temperatures the signature is weak and possibly seen after, from our observation. These simultaneous Doppler red and blue shifts in the compact but definitely extended volume (extension of order of the event diameter) shown in Figure 3 and the splitting of the strongly separated components shown on Figure 5, make us believe that this phenomenon is similar to the explosive events (EEs) which were reported by Brueckner with large velocities (Brueckner, 1980) occurring close to the surface in directions rather parallel to the underlining surface but not necessarily parallel to the LOS. Accordingly, velocities could be of higher amplitude than what is deduced from the observed

Doppler-Fizeau shifts (Figure 6). Note that the events analyzed by Brueckner were mainly described on-disk from 5 minutes duration sequences obtained during a rocket flight with few events showing only a predominantly blue shifted component, which could be the result of a peculiar positioning of the slit over a more turbulent explosive event. The components which we found split at heights which would not be seen on-disk.

After highly enhancing and sharpening our SJ images in logarithmic scale, a rather thin and long jet-like spicule was detected above the bright knot of the LEB over the background of the quiet Sun polar region (Figures 8, 9, 10, 11 and 12). It is tempting to make the connection with the LEB, knowing that the Doppler shift value is about 100 km s$^{-1}$ and the opposite wing shift similar in two split components, without taking into account the LOS effect. These values of LOS velocities could be corrected using the mean value of spicule inclination in polar regions in case we identify these components with spicules; the average tilt angle of spicules in polar areas is of order of 30 degrees or less (Tavabi, Koutchmy, and Ajabshirizadeh 2012) such that the corrected value would increase drastically by a rather large factor.

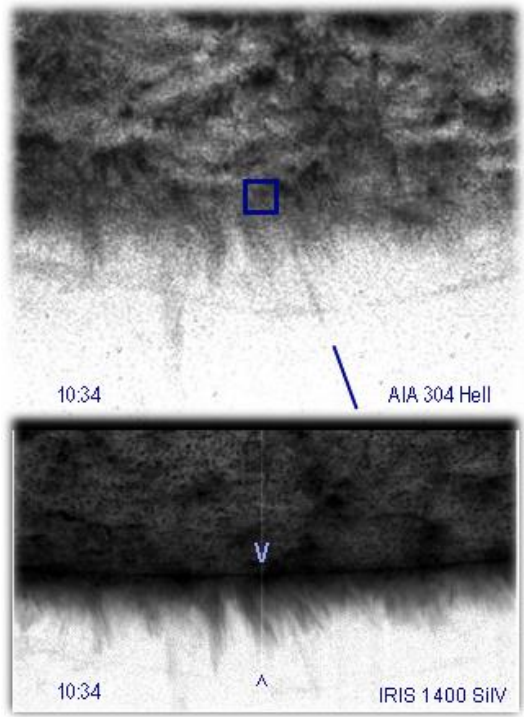

**Figure 13:** Negative partial frame from the synoptic HeII 304 Å AIA Figure 3 (top) compared to the SJ 1400 Å SiIV image taken simultaneously for showing the tiny linear thread correlated with the position of the 10:34 UT LEB. Note that in general HeII features are not well correlated with the SiIV features, just because the 304 Å resonant HeII line is much more optically thick.

Bidirectional jets in such events are accelerated in opposite directions from the reconnection sites, the first approximation X null-point reconnection geometry could imply the involvement of magnetic loops, separatrix surfaces, and spine (see *e.g.* Figure 14 for a possible scenario) for which outflow jets are actually expected to have a multidirectional structure, which indeed has been observed in the form of simultaneous blue and red Doppler shifts in the LEB taken at 10:34 UT, similar to the on-disk and larger scales events in Brueckner and Bartoe (1983), and Dere, Bartoe, and Brueckner (1989). In general, the topology of a jet resembles the geometry of field lines in the vicinity of a null point see Filippov, Golub, and Koutchmy (2009a), Filippov, Koutchmy, and Golub (2009b) and observations often show that a jet may start from the expansion of loops within a dome-like magnetic configuration.

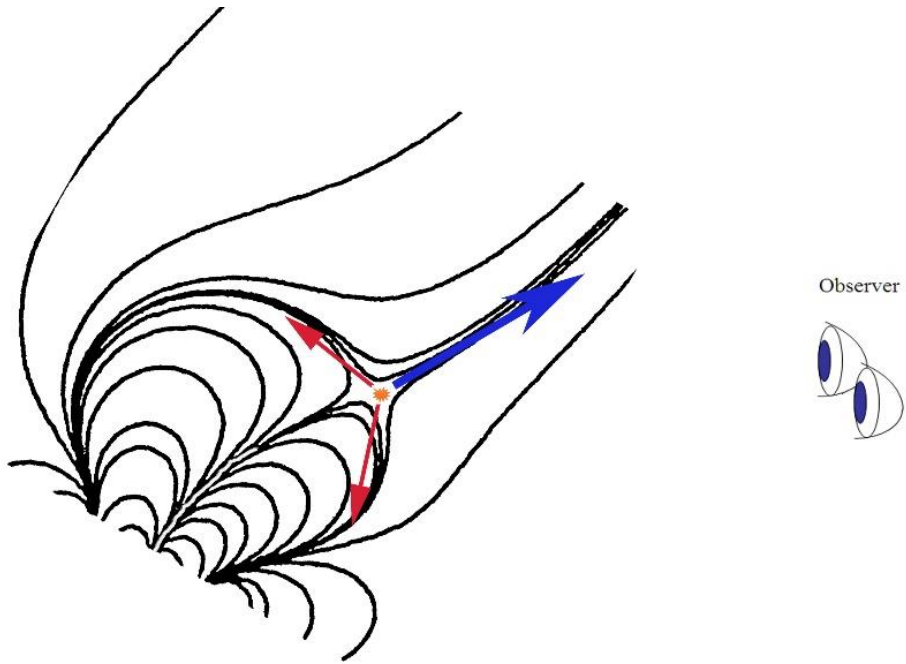

**Figure 14:** Cartoon of a possible magnetic field-line configuration near a magnetic reconnection site for the case of the 3D configuration calculated in Koutchmy et al. (1994). The figure shows three components of the velocities with two of these components (in red) in one dominant direction and the third component (in blue) in an apparent opposite direction depending on the position of the LOS and the extension of the reconnection site. The direction of motion is along the separatrix surfaces that are curved towards the feet but the spine is more linear and outwardly directed.

Another possibility is to start with a quadrupolar configuration (Filippov, Koutchmy, and Tavabi 2013) with a linear jet above resembling the Eiffel-tower configuration as was suggested for macrospicules and twin- spicules by Zirin and Cameron (1998). However with LEBs we deal with smaller energetic events at chromospheric temperatures and even a very fine thread like linear jet is seen in Si IV emissions (Figures 8, 9, 10, 11 and even in Figures 2 and 13).

Yokoyama and Shibata (1995) using *Yohkoh* soft X-ray observations showed two types of 2D magnetic configurations numerically simulated, the anemone jet type and the two-sided loop-jets type. The first one occurs when emerging flux appears in quiet regions, and two-sided loop interaction (or jet) occurs in predominantly unipolar field regions, in the horizontal direction on both sides of the emerging flux. In 3D it could resemble our naïve cartoon in Figure 14. A recent qualitative extension of this model is discussed in Singh *et al.* (2012) for events closer in size to our more energetic LEB that occur significantly above the limb. In Singh *et al.* (2012) the earlier 2D reconnection anemone model (Yokoyama and Shibata 1995) is extended to the case of a 3D twisted flux-tube reconnecting with an ambient already open field in order to explain their SOT observations of multiple knots appearing at the feet of several fine-scale jets with systematic motion. The occurrence of gradually more and more violent processing responsible for the launch of the jets was noted.

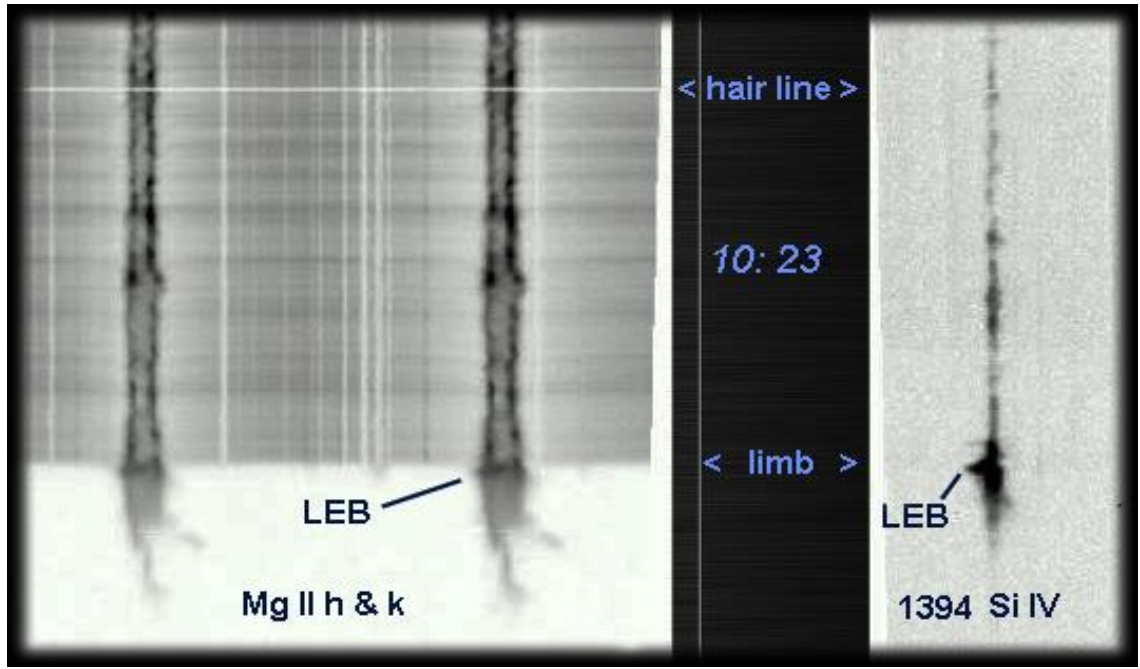

**Figure 15:** Negative display similar to Figure 4 but for another LEB observed at 10:23 UT in both the "cool" MgII lines and in the hot SiIV line 1394 Å.

Here the question "what temperatures do the LEBs reach?" arises. Would it be possible for them to reach coronal temperature? It is not possible to see the coronal counterpart of the LEB from IRIS data. The AIA (SDO) images do not have enough resolution to look at sub-arcsec events with a sufficient signal/noise ratio. In order to progress on this question we finally picked up another example of a LEB observed with the IRIS experiment at 10:24 UT. Briefly, we show in Figure 15 this LEB in different emission lines where the much hotter (De Pontieu et al., 2014; Peter et al., 2014) OIV line is seen in Figure 16 with the slit shifted by 0.35". Although the OIV line is significantly weaker in IRIS data than the SiIV lines, we clearly see the signature of the LEB, suggesting they could reach at least the TR temperature. Note also the behavior of the velocities shown by the spectral shifts of the Si IV lines for this LEB when Figures 15 and 16 are compared. From an energetic point of view, it is also worth noting that the LEBs look different from the bombs described by Peter *et al.* (2014) because the solar regions involved are completely different, the height above the photosphere is higher and finally, the OIV emission is seen in LEB and not in bombs (Peter *et al.,* 2014).

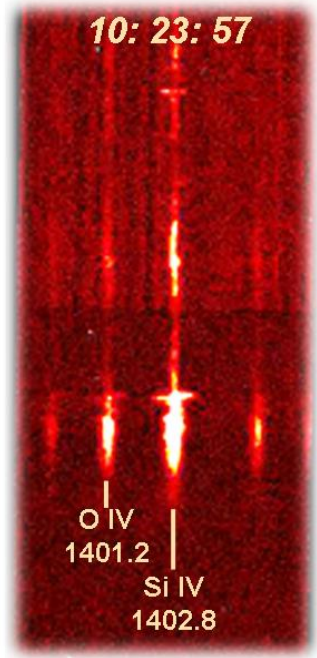

**Figure 16:** Magnified, processed and color coded positive selected part of the 1400 Å - 1404 Å spectrum of the LEB of Figure 15, just few seconds after the observation of Figure 14 with a slit position moved by 0.35". Both the emission lines of the SiIV and of the hotter but weaker OIV ions show the LEB with the signature of the large blue and red wings, above the limb shown in Figure 15 with respect to the MgII limb.

We however suggest that the velocities from the spectra seen in the TR lines are localized along directions nearly parallel to the surface with LOS effect. Note that the model proposed by Roussev *et al.* (2001) does not extend the diffusion region in the horizontal plane and would have difficulty explaining our observed split components and the occurrence of both red and blue shifted components.

## 4. Conclusions

Thanks to IRIS's outstanding resolution data where an unprecedented spectro-spatial resolution (about 1 km s$^{-1}$ precision, 1/3 arcsec in space and 10 seconds in time) was obtained in spectral rasters with simultaneous SJ images, we discovered a good example of a short duration but complex jet related to a limb event brightening (LEB) directly above the South polar limb, confirming and extending the earlier speculative interpretations given from the HRTS (Brueckner and Bartoe, 1983) photographic rocket observations. Evidence of extensions of the reconnection region in both height and azimuth (horizontal) is given with reconnection sites existing inside the 700 to 800 km side volume. At the Si IV emission temperature we observe a pair of rapidly moving separated parts like two split branches with a narrow 400 km space between them in the radial direction and another part nearby rapidly moving in the opposite direction. The signature of a LEB is also evidenced from the O IV hotter emission line. This type of spectral signature could be interpreted as evidence of 3D magnetic reconnection releasing energy at the sites where it occurs. The geometry of the magnetic region could correspond to the Eiffel-tower or to the 3D anemone configuration which eventually produces a long collimated ray that we detected on processed images as a linear thread fading out without apparent motion. The cool, dense surrounding chromosphere shows only a marginal spectral signature of these LEBs.

Our observation seems compatible with many possible scenarios that have been suggested for coronal type interactions at low plasma $\beta$, but more details need to be introduced for this type of event occurring at rather low heights, deep in the chromospheric layers (typically at 1 Mm height). The situation at these heights is more complex because the cool and dense surrounding neutral medium increases the importance of collisions and of selective radiations in emission lines during the LEB. A clear distinction is evident in the temperature "space" in the sense that almost nothing is simultaneously seen at the lower chromospheric temperature, which seems to be described here for the first time. Some limitation however occurs in our interpretation: the inherent method of observation that makes it impossible to get all the spectral data *simultaneously* because of the need for scanning the region using a very narrow slit and collecting spectra in each line to make the raster. Some ambiguity inevitably exists regarding the temporal and the spatial variations at very small scales. More statistically significant events need to be analyzed before speculating on their importance for the loading of mass into the corona and even for the coronal heating issue. Conversely, 3D models should be developed with a more extended reconnection diffusion region, made of several knots in a volume of typically 800 km size, than in the classical X-type reconnection scenario.

**Acknowledgements:** IRIS is a NASA small explorer mission developed and operated by LMSAL with mission operations executed at NASA ARC with contribution from NSC (Norway). We thank T. M. D. Pereira for his very useful online tutorial for IRIS data analysis (http://folk.uio.no/tiago/naoj2014_exercises/tutorials.html). We are indebted to Boris Filippov, Igor Veselovsky and Jean-Claude Vial for making meaningful remarks on the paper. We warmly acknowledge the work of our referee for making an extended detailed report and for many interesting suggestions and requests. This work has been supported by the Institut d'Astrophysique de Paris- CNRS and UPMC. LG was supported by a contract from Lockheed Martin to SAO.

**REFERENCES**

Aschwanden, M.: 2004, *Physics of the Solar Corona*, An Introduction, Springer, Praxis Publishing, Chichester, UK. ISBN 3-540-22321-5, First Edition, hardbound issue, 432p.

Alissandrakis, C. Zachariadis, Th., Gontikakis, C.: 2005, in Proc. 11th ESPM, ESA SP- 596

Bohlin, J. D., Vogel, S. N., Purcell, J. D., Sheeley, N. R., Jr., Tousey, R., Vanhoosier, M. E.: 1975, *Astrophys. J.* **197**, L133. DOI: 10.1086/181794.

Brueckner, G. E.: 1980, in Highlights in Astron. Reidel Publ. **5**, 557.

Brueckner, G. E., Bartoe, J.-D. F.: 1983, *Astrophys. J.* **272**, 329. DOI: 10.1086/161297.

Cirtain, J. W., et al.: 2007, Science, **318**, 1580. DOI: 10.1126/science.1147050.

Daw, A., DeLuca, E.E., Golub, L.: 1995, *Astrophys. J.* **453**, 929. DOI: 10.1086/176453.

De Pontieu, B., Title, A. M., Lemen, J. R., *et al.:* 2014, Solar Phys., **289**, 2733. DOI: 10.1007/s11207-014-0485-y.

Dere, K. P. Bartoe, J.-D., Brueckner, G.E.: 1989, Solar Phys. **123**, 41. DOI: 10.1007/BF00150011.

Filippov, B., Golub, L., Koutchmy, S.: 2009a, Solar Phys. **254**, 259. DOI: 10.1007/s11207-008-9305-6.

Filippov, B., Koutchmy, S., Golub, L.: 2009b, Geomagnetism and Aeronomy, **49**, 1109. DOI: 10.1134/S001679320908012X.

Filippov, B., Koutchmy, S., Tavabi, E.: 2013, Solar Phys. **286**, 143. DOI: 10.1007/s11207-011-9911-6.

Judge, P.G.: 2000, *Astrophys. J.* **531**, 585. DOI: 10.1086/308458.

Georgakilas, A. A., Koutchmy, S., Alissandrakis, C.F.: 1999, Astron. Astrophys. 341, 610.

Golub, L., Krieger, A. S., Silk, J. K., Timothy, A. F., Vaiana, G. S.: 1974, *Astrophys. J.* **189**, L93. DOI: 10.1086/181472.

Koutchmy, S., Stellmacher, G.: 1976, Solar phys., **49**, 253. DOI: 10.1007/BF00162449.

Koutchmy, S., Loucif, M.: 1991, in "Mechanism of Chrom. and Coronal Heating", Conf. Heidelberg, Ulmschneider, Priest and Rosner Ed., Springer- Verlag, 152.

Koutchmy, S., Koutvitsky, V.A., Molodensky, M.M., Soloviev, L.S., Koutchmy, O.: 1994, Space Sci. Rev., **70**, 283. DOI: 10.1007/BF00777881.

Koutchmy, S. Hara, H. Suematsu, Y., Reardon, K.: 1997, Astron. Astrophys. **320**, L33.

Moore, R. L., Cirtain, J. W., Sterling, A. C. and Falconer, D. A.: 2010, *Astrophys. J.* **720**, 757. DOI: 10.1088/0004-637X/720/1/757.

Moore, R. L., Tang, F., Bohlin, J. D., Golub, L.: 1977, *Astrophys. J.* **218**, 286. DOI: 10.1086/155681.

Pasachoff, J. M., Jacobson, W. A., Sterling, A. C.: 2009, Solar Phys. **260**, 59. DOI: 10.1007/s11207-009-9430-x.

Peter, H., Tian, H., Curdt, W. *et al.*: 2014, Science, **346**, 315. DOI: 10.1126/science.1255726.

Innes, D. E., Inhester, B., Axford, W. I., Wilhelm, K.: 1997, Nature, **386**, 811. DOI: 10.1038/386811a0.

Roussev, I., Galsgaard, K., Erdelyi, R., Doyle, J.G.: 2001, Astron. Astrophys**. 375**, 228. DOI: 10.1051/0004-6361:20010765.

Sekse, D. H., Rouppe van der Voort, L., De Pontieu, B.: 2012, *Astrophys. J.* **752**, 108. DOI: 10.1088/0004-637X/752/2/108.